\documentclass[a4paper,fleqn,usenatbib]{mnras}

\usepackage[T1]{fontenc}
\usepackage{ae,aecompl}

\usepackage{graphicx}	% Including figure files
\usepackage{amsmath}	% Advanced maths commands
\usepackage{amssymb}	% Extra maths symbols
\usepackage{threeparttable} 

\title[Multiple Outbursts of MAXI~J1957+032]{Swift and SALT observations of the Multiple Outbursts of MAXI~J1957+032}

\author[Mata S\'{a}nchez et al.] {D. Mata S\'{a}nchez$^{1,2}$\thanks{E-mail:dmata@iac.es}, P.A. Charles$^{1, 3}$, M. Armas Padilla$^{1,2}$, D.A.H. Buckley$^4$,
\newauthor
G.L. Israel$^5$, M. Linares$^{1,2}$, T. Mu\~{n}oz-Darias$^{1,2}$
\\
$^{1}$Instituto de Astrof\'{i}sica de Canarias (IAC), E-38205 La Laguna, Tenerife, Spain\\
$^{2}$Departamento de astrof\'isica, Univ. de La Laguna, E-38206 La Laguna, Tenerife, Spain\\
$^{3}$Department of Physics \& Astronomy, University of Southampton, Southampton SO17 1BJ, UK\\
$^{4}$South African Astronomical Observatory, PO Box 9, 7935 Observatory, Cape Town, South Africa\\
$^{5}$INAR-Osservatorio Astronomico di Roma, via Frascati 33, I-00040 Monteporzio Catone, Italy
}

\date{Accepted 2017 February 21. Received 2017 February 20 ; in original form 2017 January 12}

\pubyear{2017}
\begin{document}
\label{firstpage}
\pagerange{\pageref{firstpage}--\pageref{lastpage}}
\maketitle

\begin{abstract}
The new recurrent X-ray transient MAXI~J1957+032 has had four X-ray outbursts within 16 months, all very briefly detected (lasting $<$5 days). During the most recent event (Sep/Oct 2016), we obtained with SALT the first optical spectrum of the transient counterpart, showing the classic blue continuum of an X-ray irradiated disc in an LMXB and no other features. At high Galactic latitude below the plane (-13$^o$) reddening is low but there is no quiescent counterpart visible on any of the existing sky surveys, nor any other known X-ray source in the region. \textit{Swift} monitoring of 3 of the 4 events is presented, showing rapidly fading X-ray outbursts together with significant UVOT detections in the UV (W1,M2,W2), U and B bands. The optical properties are most like those of the short-period LMXBs, which, combined with the softening witnessed during the decay to quiescence would place the system at $d <13 \, \rm  kpc$. The short duration and short recurrence time of the outbursts are reminiscent of the AMXPs, which exhibit peak luminosities of $\sim 1\% \, \rm L_{Edd}$. Assuming this peak luminosity would place MAXI~J1957+032 at a distance of $d\sim 5-6\, \rm  kpc$.
%we propose that MAXI~J1957+032 could be a new member of this class, with a peak luminosity of $\sim 1\% \, \rm L_{Edd}$ at a distance of $\sim \rm 5-6\, kpc$.
\end{abstract}


%%%%%%%%%%%%%%%%%%%%%%%%%%%%%%%%%%%%%%%%%%%%%%%%%%

%%%%%%%%%%%%%%%%% BODY OF PAPER %%%%%%%%%%%%%%%%%%

\section{Introduction}

The highly variable X-ray sky is populated with X-ray transients (XRTs). These are usually classified as objects typically in quiescence with X-ray luminosities well below $10^{33} \rm erg\, s^{-1}$, which increase by several orders of magnitude during outbursts. Within our Galaxy, luminous X-ray transient behaviour is demonstrated by both high-mass and low-mass X-ray binaries (HMXBs and LMXBs respectively), where the former are predominantly BeX systems involving neutron stars in long (tens of days or more) eccentric orbits around a rapidly rotating Be star (e.g. \citealt{2011Ap&SS.332....1R}). As we will show, the properties of J1957 are consistent with that of LMXB transients, which contain either a black hole (BH) or neutron star (NS) in a close (typically $P_{\rm orb} \sim$ hours) circular orbit with a low-mass, evolved donor (\citealt{Remillard2006}, \citealt{Liu2007}).  

Since the demise of the \textit{RXTE} all-sky monitor (ASM) in 2012, the X-ray sky monitoring task has been shared between the \textit{INTEGRAL}, \textit{Swift} and \textit{MAXI} missions. Thanks to their wide field and all-sky monitoring programs (\citealt{Kuulkers2007}, \citealt{Krimm2013} and \citealt{2009PASJ...61..999M}, respectively), the last decade has seen the emergence of a number of new XRTs (Swift~J1753.5-0127, MAXI~J1659-152, Swift~J1357.2-093313, MAXI~J1305-704) at high Galactic latitude and subsequently found to have short periods ($\leq 5\, \rm h$).  Combined with the already well-known XRTs GRO~J0422+32 and XTE~J1118+480, this suggested the existence of a possible sub-group of LMXB XRTs, whose high latitude (and consequent low extinction) made them important targets for multi-wavelength studies (see e.g. \citealt{2013MNRAS.433..740S}).  Furthermore, since many of these XRTs are too faint in quiescence for current telescopes (especially the short period systems which tend to have the lowest mass donors), it is very important to undertake rapid follow-up optical spectroscopy while the X-ray outburst is underway, as it can then be possible to exploit the X-ray irradiation of the donor to provide constraints on the motion of the donor (see e.g. \citealt{2013ASPC..470..263C}).

The high-latitude XRT MAXI~J1957+032 (hereafter ``J1957'') was discovered in 2015, independently by both \textit{MAXI} \citep{2015ATel.7504....1N} and \textit{INTEGRAL} (\citealt{2015ATel.7506....1C}, who designated the source IGR J19566+0326), and optically identified \citep*{2015ATel.7524....1R} with a faint $r'\sim20$ star, there being no counterpart on sky-survey images \citep{Chakrabarty2016}. However, that  and the subsequent two outbursts over the next year (\citealt{2015ATel.8143....1S}, \citealt{2016ATel.8529....1T}\footnote{Note that the 2-20 keV fluxes quoted for Jan 7 and 8 are incorrect. They should have been 0.066 and 0.071 $\rm ph\, cm^{-2} s^{-1}$ respectively}), decayed rapidly (within a matter of days), making extensive follow-up study almost impossible.  Nevertheless, \citet{2015ATel.8197....1M} described a low-resolution ($\sim$17\AA) optical spectrum of J1957 taken during the Oct 2015 outburst (where \citealt{2015ATel.8149....1G} had reported R=18.3 four days after the start of the outburst), which contained only absorption features consistent with an F/G-type star, and no emission lines. This was surprising, as virtually all LMXB XRTs in outburst have optical spectra that are dominated entirely by the X-ray irradiated accretion disc, producing a strong blue continuum on which is usually superposed strong Balmer, HeII and Bowen emission features (see e.g. \citealt{2006csxs.book..215C}). 

Furthermore, to have 3 X-ray outbursts within a year is unusual amongst the class of LMXB XRTs; so, when a new, brighter X-ray outburst of MAXI~J1957+032 was reported in Sep 2016 \citep{2016ATel.9565....1N}, we immediately initiated our SALT Transients ToO program \citep {2016ATel.9649....1B} in order to obtain further optical spectra of this intriguing object. Those data, together with the associated \textit{Swift} monitoring of J1957, are the subject of this paper.

\section{Observations}
\label{observations}
Due to the short duration and faintness of J1957 outbursts, both \textit{MAXI}/GSC (2-20 keV) and \textit{Swift}/BAT (15-50 keV) light curves do not exhibit any remarkable feature (e.g. rebrightening) other than the reported starting times of the 4 events. These correspond to 57153.58 (2015-05-11, epoch 1), 57301.52 (2015-10-06, epoch 2), 57394.35 (2016-01-07, epoch 3) and 57660.35 (2016-09-29, epoch 4); 3 of them marked in Fig.~\ref{fig: zoom}.

\subsection{Swift XRT}

\subsubsection{Spectral analysis}

%%%%%%%%%%Table OBS %%%%%%%%%%%%
\begin{table*}
\centering
\caption{\textit{Swift} observations log and spectral results for J1957 using a PHABS*PO model. We assumed a constant $N_{\rm  H}$ of $1.7\times10^{21}~\rm cm^{-2}$, inferred from the highest signal-to-noise spectrum (00033770017).}
\begin{threeparttable}
\begin{tabular}{l c c c c r c}

%\multicolumn{3}{c}{Observations}  & \multicolumn{2}{c}{Date} & \multicolumn{2}{c}{XIS}& HXD-PIN\\

\hline
 \multicolumn{1}{c}{ID} &  MJD          & Mode &  Exposure &    $\Gamma$   &    \multicolumn{1}{c}{ $unabs,F_{\rm x}$ (0.5-10~keV) }  & $\chi^{2}_{\nu}$ (dof)\\   
             			&         &		  &[ks]		  &				  &		\multicolumn{1}{c}{		[$\mathrm{erg~cm}^{-2}~\mathrm{s}^{-1}$	]	}	&				\\
\hline
00033770001 & 57155.96 & PC   &    2.9   &   2.13 $\pm$   0.07   & (1.55$\pm$ 0.08) $\times10^{-11}$   	 &0.97 (167) \\ 
00033776001 & 57156.42 & PC   &    0.44 &   1.5 $\pm$    0.6    & (8.5$^{+6.2}_{-3.7}$ ) $\times10^{-12}$ & 0.36 (5)   \\
00033777001 & 57156.43 & PC   &    0.45 &   2.5  $\pm$   0.8    & (5.2$^{+2.6}_{-1.6}$) $\times10^{-12}$  & 0.91 (6)    \\
00033770009 & 57304.67 & PC   &    0.96 &   2.0  $\pm$   0.1    & (1.61$\pm$ 0.16) $\times10^{-11}$       & 1.11 (67)  \\
00033770010 & 57305.61 & PC   &    0.98 &   2.6  $\pm$   0.9    & (1.5$^{+0.7}_{-0.5}$) $\times10^{-12}$  & 1.54 (5)    \\
00033770017 & 57660.69 & WT   &    0.97 &   1.90 $\pm$   0.02   & (9.13$\pm$  0.13) $\times10^{-10}$      & 1.02 (364) $^{b}$ \\                 
00033770018$^{a}$ & 57661.09 & WT   &    0.77 &   1.92 $\pm$   0.02   & (7.09$\pm$ 0.12) $\times10^{-10}$       & 1.04 (304) $^{b}$     \\
00033770018-1$^{a}$ & 57661.09 & WT & 	  0.45 &   1.92 $\pm$   0.03   & (8.65$\pm$  0.18) $\times10^{-10}$      & 1.18 (252) $^{b}$\\
00033770018-2$^{a}$ & 57661.62 & WT &    0.33 &   1.93 $\pm$   0.04   & (5.53$\pm$ 0.18)  $\times10^{-10}$      & 0.89 (153) $^{b}$\\
00033770019 & 57662.75 & PC   &    0.37 &   1.89 $\pm$    0.2   & (1.58  $\pm$ 0.22)$\times10^{-10}$      & 0.84 (80) \\                   
00033770020 & 57663.08 & PC   &    1.7 &   1.94 $\pm$0.07& (2.80 $\pm$ 0.17) $\times10^{-11}$      &  1.22 (164) \\    
00033770021 & 57664.14 & PC   &    1.7	&   2.3  $\pm$   0.3    & (3.03$\pm$ 0.49) $\times10^{-12}$       &   0.97 (27) \\                  

\hline
\end{tabular}
\begin{tablenotes}
\item[a]{Observation 00033770018 was divided in two sub-events of detections $\sim$13h apart (see Section 2.1.1). }
\item[b]{Spectra with a minimum of 20 photons per bin. The rest are grouped with a minimum of 5 photons per bin (see Section 2.1.1).}
\end{tablenotes}
\label{tab:results}
\end{threeparttable}
\end{table*}

\begin{figure}
\scalebox{1.0}{\includegraphics[keepaspectratio, trim=0.5cm 0.cm 2cm 0.5cm, width=\columnwidth]{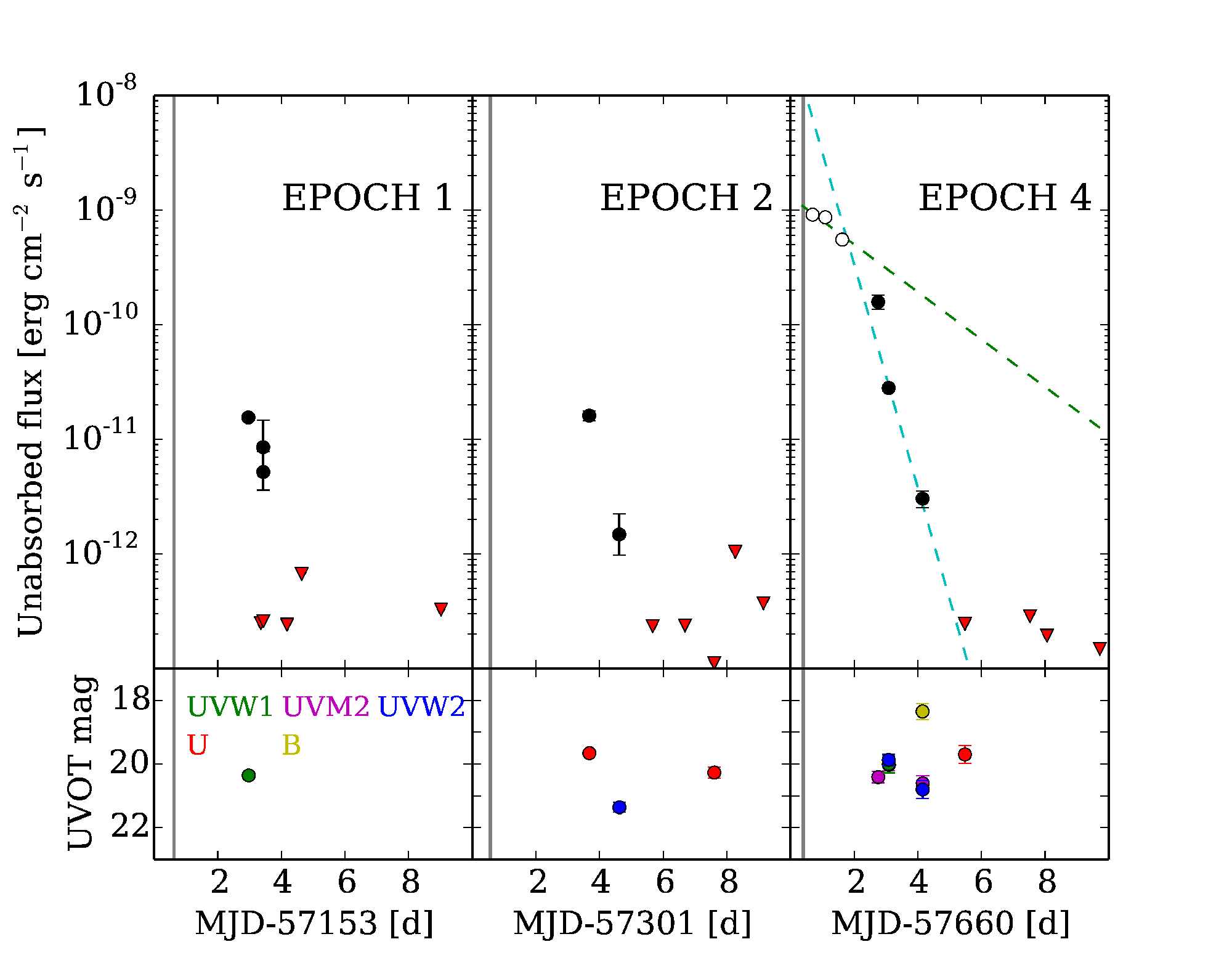}}
    \caption{Upper panel: Detailed \textit{Swift}/XRT (0.5--10~keV) monitoring of J1957 events 1, 2 and 4. The most detailed coverage is of event 4, where we also plot exponential decays of e-folding timescales $2.1\pm 0.9$ and $0.44\pm 0.15$ days to the slow and fast declines respectively. Empty circles refer to WT mode and filled circles to PC mode observations. The start times of the covered outbursts are marked by vertical dashed lines. Lower panel: \textit{Swift}/UVOT (B, U, UW1, UM2 and UW2 filters) follow-up pointings are shown.}
    \label{fig: zoom}
\end{figure}

\begin{figure}
\scalebox{1.0}{\includegraphics[keepaspectratio, trim=0.5cm 0.cm 0.3cm 0.cm, width=\columnwidth]{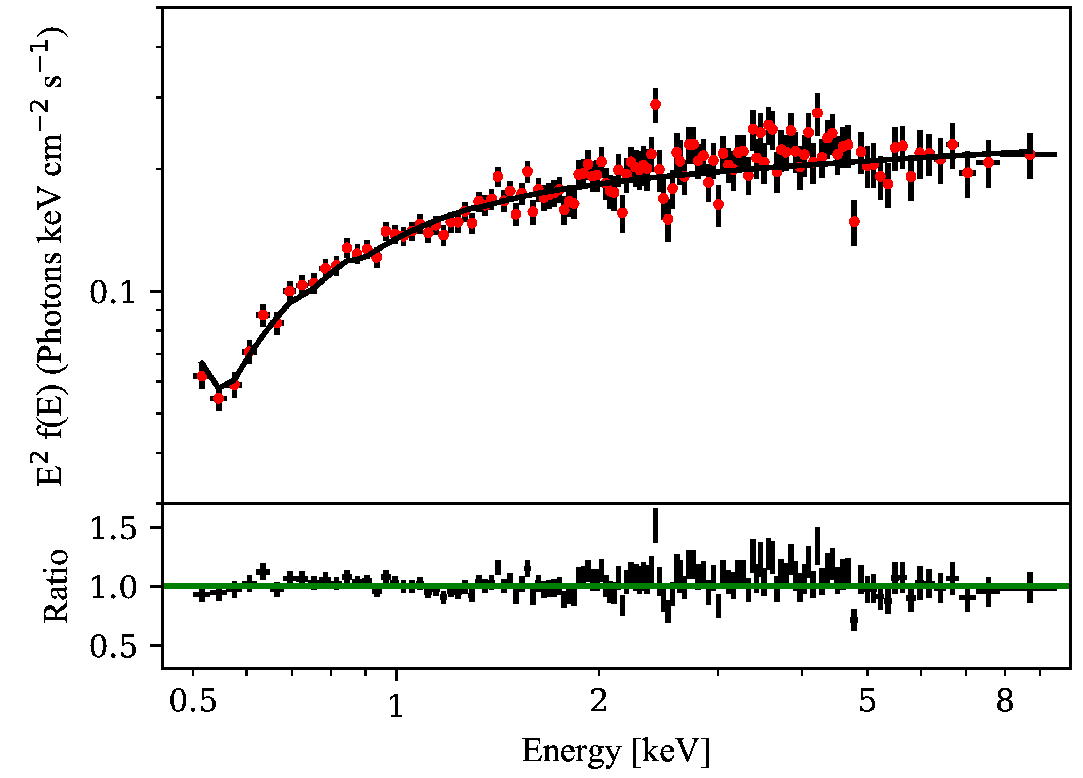}}
    \caption{Unfolded spectra of observation 00033770017 (57660.69, 2016-09-29). The solid line represents the best-fitting using a PHABS*POWERLAW model. The sub-panel shows the data-to-model ratio. Data have been re-binned in XSPEC for clarity.}
    \label{fig: xspectrum}
\end{figure}

Following the outburst alerts the X-ray Telescope (XRT; \citealt{Burrows2005}) on board \textit{Swift} \citep{Gehrels2004} covered 3 of the 4 reported events. 
A total of 28 observations (2 in window timing [WT] mode and the rest in photon counting [PC] mode) were performed but, due to the outburst rapid decline, J1957 is detected in only 10 of these observations. We processed the data making use of the {\ttfamily HEASoft} v.6.18 software. The data reduction was carried out running the {\ttfamily xrtpipeline} task in which standard event grades of 0-12 and 0-2 were selected for the PC and WT mode observations, respectively. For each observation the 0.5--10~keV spectrum, light curve and image were obtained using {\ttfamily Xselect}. We used a circular region of $\sim 40$~arcsec radius centred at the source position (the inner 10~arcsec were excluded for 00033770019 observation, 57662.75, affected by pile-up). For the PC observations, three circular regions of similar size and shape, positioned on an empty sky region, was used for the background. On the other hand, an annulus centred on the source with $\sim$82 pixels for the inner radius and $\sim$118 pixels for the outer radius was used for the background of the WT observations. We created exposure maps and ancillary response files following the standard \textit{Swift} analysis threads\footnote{\url{http://www.swift.ac.uk/analysis/xrt/}}, and we acquired the last version of the response matrix files from the High Energy Astrophysics Science Archive Research Center (\textsc{HEASARC}) calibration database (CALDB). Finally, we grouped the spectra to have a minimum of 20 photons per bin to be able to consistently use the $\chi^{2}$. However, due to the low number of counts collected during various observations, we fitted these data using both C-statistic and $\chi^{2}$ but grouped them with a minimum of 5 photons per bin (see Table \ref{tab:results}). The results using both methods were consistent with each other (e.g; \citealt{Wijnands2002b}; \citealt{ArmasPadilla2013a}).
We fitted the spectra using \textsc{xspec} (v.12.9, \citealt{Arnaud1996}). All observations were well-fit with a simple power law (PL) plus absorbing column (PHABS in \textsc{xspec}) assuming the cross-sections of \citet{Verner1996} and the abundances of \citet{Wilms2000}. We assumed a constant hydrogen equivalent column density of $N_{\rm  H}=1.7\times10^{21}~\rm cm^{-2}$, inferred from our highest signal-to-noise spectrum (Fig.~\ref{fig: xspectrum}) and consistent with the low reddening expected for that direction (see next section). The photon-index of each spectrum's fit varies from $ 1.90\pm 0.02 $ to $2.34^{+0.27}_ {-0.25}$ (from highest to lowest luminosity). The unabsorbed fluxes are reported in Fig.~\ref{fig: zoom}. We note that observation 00033770018 (57661.09) consists of two $\rm \sim 350\, \rm s$-long groups of detections $\sim13\, \rm h$ apart (similar to the gap between the first group and the previous observation, $\sim14\, \rm h$), so we decided to plot two independent points in the light curve.

For the observations where J1957 is not detected, we determined the 95 per cent confidence upper limit count rates using the prescription given by Gehrels (1986). We estimated the corresponding unabsorbed flux upper limits using the WebPIMMS \textsc{HEASARC} tool. We co-added 10 Swift/XRT (PC mode) observations (11.7 ksec), in order to constrain the quiescent luminosity of MAXI~J1957+032. We can thereby place a 95\% confidence upper limit on the 0.5--10~keV quiescent unabsorbed flux of $ 1-3\times 10^{-13} \, \rm erg\, cm^{-2} s^{-1}$ (assuming $ N_{\rm H}= 1.7\times 10^{21} \rm cm^{-2}$ and a power law photon index in the 1--3 range). Comparing with the epoch 4 detection at a flux of $ \sim10^{-9}\, \rm erg\, cm^{-2} s^{-1}$, this reveals that the X-ray luminosity dropped by at least 4 orders of magnitude in just about a week.

\subsubsection{Timing Analysis}

We searched for pulsations and quasi-periodic
oscillations (QPOs) in the three \textit{Swift}/XRT observations with
windowed timing (WT) mode data (1.78 ms time resolution). These were taken on 57160 (2015-05-18, 1.3-ksec exposure), 57660 (2016-09-29, 1 ksec) and 57661 (2016-09-30, 0.8 ksec). We used 30-s and 120-s-long Fast Fourier Transforms (FFTs), including energies between
1 and 10 keV, keeping the original time resolution and
inspecting individual FFTs as well as observation averages. We detect
neither coherent pulsations nor aperiodic variability in the
0.1--281~Hz frequency range. The most constraining upper limits come from the 57660
observation, when the average 1--10~keV count rate was
$\sim$16.5~c/s. Following \citet{1994ApJ...435..362V}, we
estimate upper limits on the pulsed fraction of a sinusoidal
signal. We find upper limits on the 0.1--281~Hz pulsed fraction
between 2.2\% and 5.4\% (95\% confidence level), depending on the FFT
length and exact frequency range. We note that, considering the pulsed
fractions and spin frequencies of AMXPs (e.g. \citealt{2012arXiv1206.2727P}, and references therein), this result does not exclude the presence of similar pulsations in J1957.

\begin{figure*}
\scalebox{0.5}{\includegraphics[width=7cm]{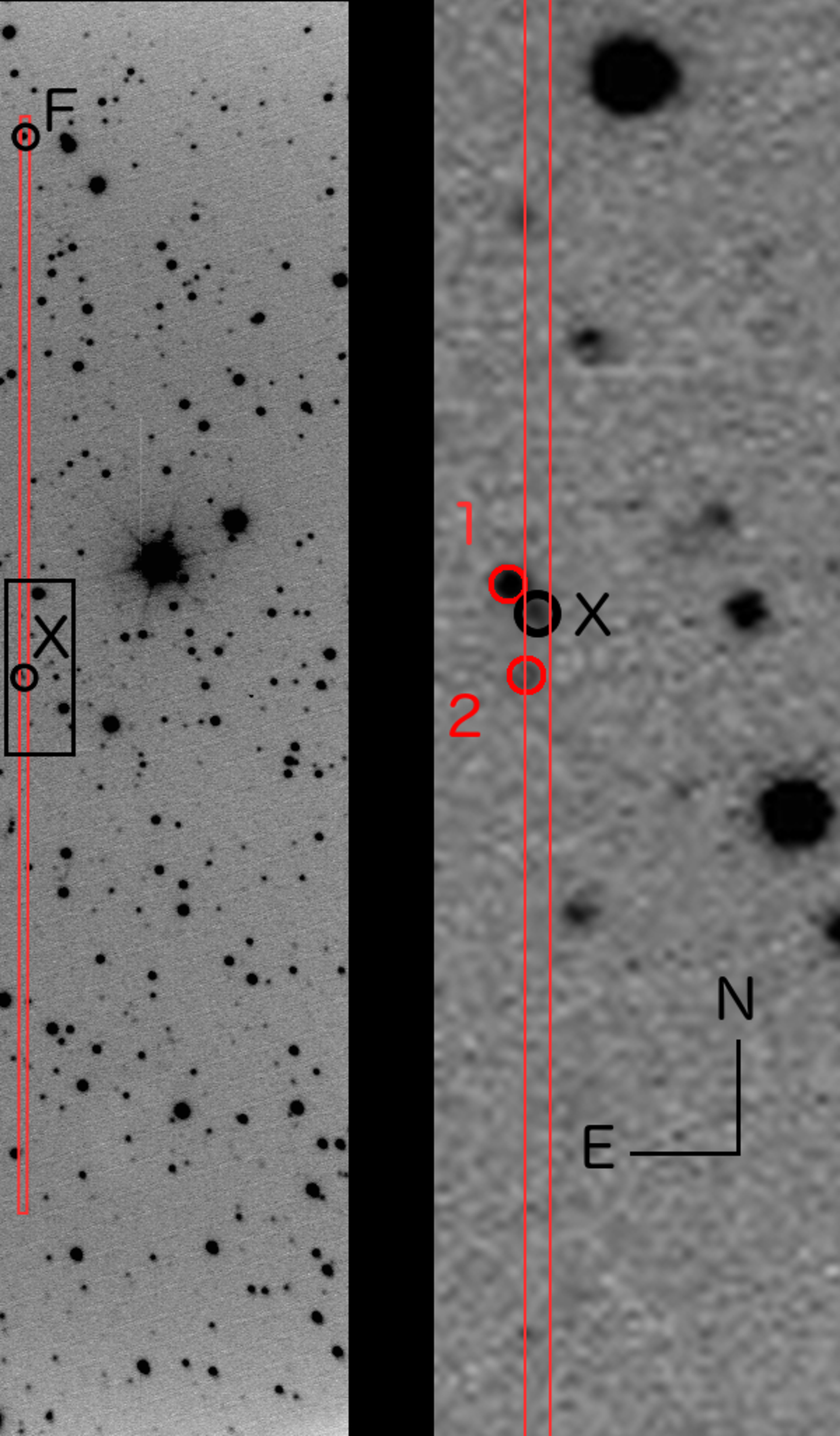}\includegraphics[width=32cm]{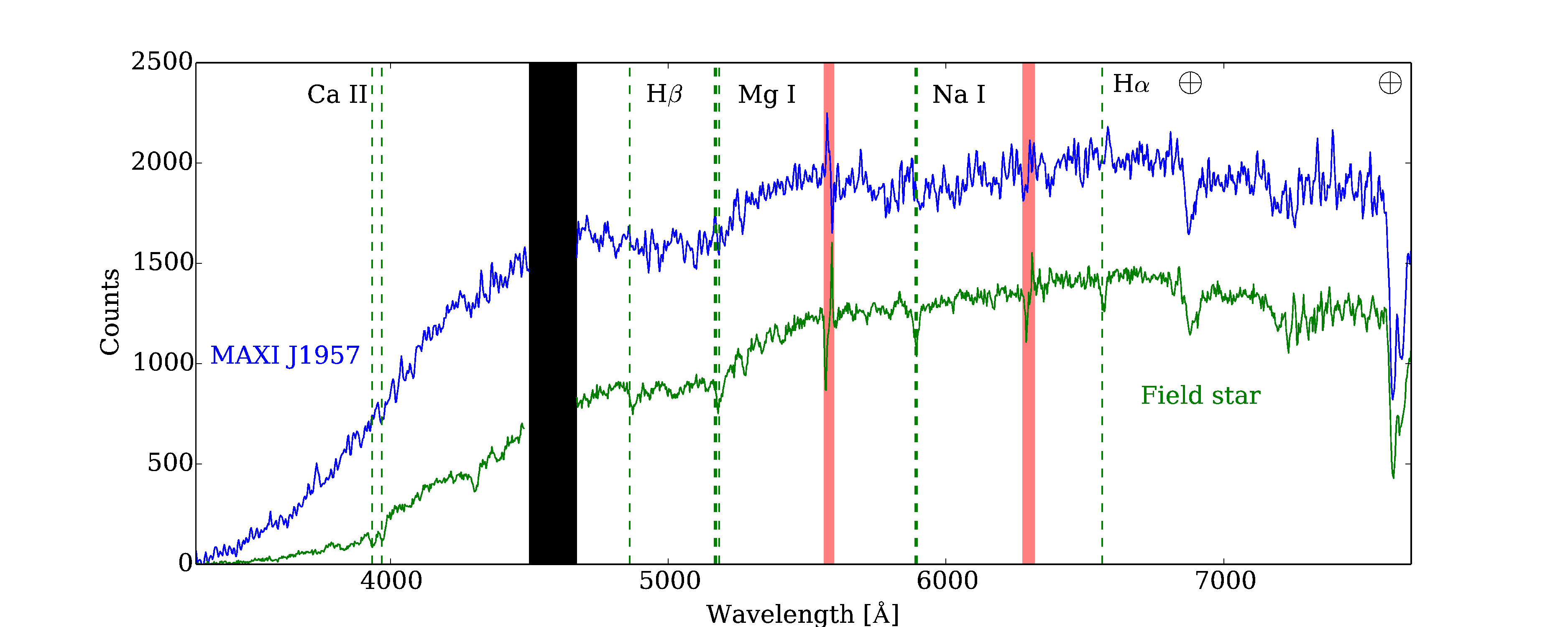}}
    \caption{Left panel: Part of the SALTICAM r' band image (left) of the region of J1957 (marked with ``X'' and a black circle) showing the orientation of the RSS long slit (red box) and the field star ``F'' that was included.  The black box is blown up (right) so as to show the location of J1957 and its nearby stars marked ``1'' and ``2'', circled in red. Right panel: RSS low-resolution spectra of J1957 (upper) and the simultaneously observed brighter field star (``F'') whose flux has been reduced by x7 to facilitate comparison. ``F'' is a late-G star in the USNO catalogue (see text) and the key spectral features are marked as green-dashed lines. Telluric features are marked with the symbol $\oplus$ and sky subtraction residuals as red-shadowed bands. There are no cosmic features in J1957, but it is noticeably bluer than ``F''. Color figures are available in the online version.}
    \label{fig: salt}
\end{figure*}

\subsection{\textit{Swift}/UVOT}

There were also optical/UV observations of J1957 associated with many of the XRT pointings, and these were accessed from the Swift online archive. We used the \textit{uvotsource} software (released as part of the HEASOFT package) to extract fluxes for J1957 using a 5~arcsec aperture centered at the position reported in  \citet{2015ATel.7524....1R}. J1957 was not detected in the V band, our best constraint being V>17.85 (MJD 57665.5). At different times it was seen in all other bands, with the brightest detections being B=$18.5\pm 0.3$ (MJD 57664.1), U=$18.7\pm 0.3$ (MJD 57665.5), UVW1=$18.5\pm 0.3$, UVM2=$18.3\pm 0.2$ and UVW2=$18.1\pm 0.2$ (MJD 57663.1). All values are Vega magnitude and were obtained during epoch 4. Using the relation  $N_{\rm H} = 2.0\pm 0.5 \times 10^{21}~\rm{cm}^{-2}\, A_{\rm V}$ \citep{Watson2011} and the Galactic hydrogen column density $N_{\rm H}=0.95\times10^{21}~\rm cm^{-2}$ (which yields $ E_ {\rm B-V}=0.14$, \citealt{Willingale2013}); we obtain $A_{\rm V}\sim  0.5$ and $A_{\rm B}\sim 0.6$. This implies the brightest B-band detection corresponds to an intrinsic value of $\rm B=17.9\pm 0.3$. We note that using a 5~arcsec aperture includes star ``1'' (see Fig. \ref{fig: salt}), classified as an A-F star (R = 18.3). It makes a small contribution in the B band (B $>$ 19), has no effect in the UV-bands, but affects V-band observations which will be disregarded.

\subsection{SALT Imaging and Spectroscopy}

We obtained optical imaging of J1957 with the Southern African Large Telescope (SALT, \citealt{2006SPIE.6267E..0ZB}) on 57662.76 (2016-10-01, epoch 4), 2.4 d after the initial X-ray outburst report \citep{2016ATel.9565....1N}.  We used SALTICAM \citep{2006MNRAS.372..151O} to obtain a 60 s r' band image (Fig.~\ref{fig: salt}), which showed the previously identified optical counterpart \citep{2015ATel.8149....1G, 2015ATel.8197....1M} to be at r' = 18.5$\pm$0.15, similar to the R=18.27 reported 4.2d after the October 2015 outburst \citep{2015ATel.8149....1G}, and brighter than the r' = 20.03 measured 3.7  after the 2015 May outburst \citep{2015ATel.7524....1R}.

With this clear indication of renewed optical activity, we then immediately followed this with a 1000s exposure spectrum using the SALT Robert Stobie Spectrograph (RSS).  This was taken with the 300 l/mm grating (mean resolving power of 320) covering 360-770nm and through a 1.5~arcsec slit (the PSF was 1.3~arcsec FWHM), oriented as shown in Fig.~\ref{fig: salt} so as to exclude light from the nearby star ``1'' (2.0~arcsec to the NE of J1957).  These data, together with associated arcs and bias frames were processed and reduced with the standard SALT pipeline \citep{2010SPIE.7737E..25C}, followed by cosmic-ray removal using the IRAF\footnote{IRAF is distributed by National Optical Astronomy Observatories, operated by the Association of Universities for Research in Astronomy, Inc., under contract with the National Science Foundation.} task ``L. A. Cosmic'' \citep{2001PASP..113.1420V} on the 2D image.  We then extracted the spectrum of J1957 and other stars within the slit using standard IRAF routines. Wavelength calibration was performed using arc spectra with the MOLLY routines {\ttfamily arc} and {\ttfamily acal}, and the results are in Fig.~\ref{fig: salt}.

The spectrum of J1957 is essentially featureless (apart from telluric lines), notably so in comparison with field star ``F'', which was simultaneously in the slit. This star is USNO-B1.0 0935-0543890, of observed spectral type G5-K0 and with B=18.3, B-R=1.2, which we have used with our spectra to estimate B-V=0.3 for J1957 (significantly bluer than ``F'' in our spectrum for which we obtain B-V=0.7). X-ray active LMXBs have optical spectra dominated by the X-ray irradiated disc, and often have strong Balmer and HeII lines in their spectra. However, the more compact systems, especially those seen at high inclination, have much weaker emission lines, often almost completely undetectable (e.g. \citealt{2016A&A...587A.102B}). Indeed, ultracompact systems accretion discs should be depleted of hydrogen, as the companion star is itself degenerate (see \citealt{2010NewAR..54...87N}). We constrain the equivalent width of the H$\alpha$ line to be lower than $3\, \textrm{\AA}$ at $3\sigma$ confidence level (assuming a conservative $\rm{FWHM}<4400\, \textrm{\AA}$ determined by the widest H$\alpha$ profile measured so far in a LMXB, see \citealt{MataSanchez2015b}).

As already reported \citep{2016ATel.9649....1B}, our spectrum contrasts with the A-F star absorption features reported by \citet{2015ATel.8197....1M} using the Nordic Optical Telescope (and ALFOSC) at a similar resolution 8.2 d after the epoch 2 outburst (2015 October). We re-analysed the ALFOSC data of this epoch, and from a 250 s image taken immediately after the spectrum, obtained R=21.4$\pm$0.2 for J1957. This led to the realisation that the reported NOT spectrum was not of J1957, but rather of star ``1'', 2.0~arcsec NE of J1957. We measured star ``1'' to have r' = $18.0\pm 0.15$ and R = $18.3\pm 0.1$, respectively, from the SALTICAM and ALFOSC images taken a year apart, and is consistent with the fluxes measured in the NOT spectrum. All of the photometry reported here used the same reference star, USNO B1.0 0934-0560330.

\section{Discussion}

Having resolved the question of the spectroscopic identification of J1957, and demonstrated that its optical spectrum is consistent with other LMXB XRTs in their active phase, the \textit{Swift}/XRT light curve highlighted the extremely rapid decay of the outburst (and which Fig.~\ref{fig: zoom} shows was likely the case for all the outbursts).  The essential properties of this object are straightforward to summarise: (i) outburst light curves that increase rapidly (within a \textit{MAXI} orbit we change from non-detection to the peak luminosity) and decay steeply, lasting $<5\, \rm d$ in total; (ii) no detected rapid variability or type I X-ray bursts during their bright phase; (iii) have an associated optical brightening from quiescence of $\geq$3 mags; (iv) hard X-ray spectrum (not that of a disc black-body). With these properties at hand, we can rule out the following classifications for J1957:

\begin{itemize}
 \item HMXB/BeX system: the outbursts of J1957 have already been noted to be non-periodic \citep{2016ATel.8529....1T}, and the large range of optical variability (combined with our SALT spectrum) from quiescence rules out the possibility of an early-type donor.;
 \item superburster: the longest ``superburst'' is $\sim 0.3\, \rm d$ \citep{2011arXiv1102.3345I}, much less than the 5 days of each J1957 outburst, and there would be enormous theoretical difficulties in producing sufficient nuclear fuel for such a superburst;
  
\end{itemize}

\subsection{LMXB/XRT: comparison with Galactic Centre Region X-ray Transients}
In spite of the extensive \textit{RXTE}/ASM database of XRT light-curves (see e.g. \citealt{2015ApJ...805...87Y}) there are no BH LMXB/XRT which display properties comparable to those of J1957. Within that same database, only the NS system MXB~1730-335 (better known as the ``Rapid Burster'') has a long-term light-curve similar to that of J1957.  Indeed, in the latter phases of the \textit{RXTE} mission, MXB~1730-335 underwent 6 outbursts in barely 2 years, each of which had a bright phase of usually $<$10 days. However, in its bright phase, this object emits most of its flux in the form of very rapid, type II X-ray bursts (e.g. \citealt{2003PASJ...55..827M}) that last only for 10s or so. J1957 exhibits no such variations.

More recently, \citet{Degenaar2015} reported a 9 years \textit{Swift} X-ray monitoring of the Galactic centre. During this campaign, various short outbursts ($\sim 15$ days long) were detected  peaking at $L_{X}\sim 10^{35} \rm erg\,s^{-1}$, both from very faint as well as brighter LMXBs. However, none of these have a recurrence time as short as J1957.

Perhaps the most comprehensive monitoring campaign on Galactic X-ray transients took place in the period 1996-2000 when the BeppoSAX Wide Field Cameras (which had a 40$^{\rm o}$ field of view) observed the Galactic Centre region nine times for a total exposure of $\sim 4\, \rm Ms$.  As summarised by \citet{2001ESASP.459..463I}, the combination of the BeppoSAX WFC observations and the \textit{RXTE}/ASM database led to the discovery of 31 LMXB XRTs that were active during this time interval.  From this study (see their Table 2) there are 4 XRTs whose duration and peak X-ray flux are comparable to J1957. These are KS~1741-293, SAX~J1748.9-2021, SAX~J1750.8-2900 and SAX~J1810.8-2609, all of whose outbursts lasted less than 10 days.  However, subsequent \textit{INTEGRAL} observations of the former \citep{2007MNRAS.380..615D} showed more extended periods of activity, and the latter two \citep{1999ApJ...523L..45N, 2000ApJ...536..891N} both displayed Type I X-ray bursts with close to Eddington-limited properties. Despite J1957 not exhibiting any of those features, we note that either we could have missed the ignition of a hypothetical burst or maybe its outbursts are too short to trigger it. The remaining system SAX~J1748.9-2021 is an AMXP (accreting millisecond X-ray pulsar) spinning at 442Hz in an 8.8hr binary, and while it has recurrence times of order a year, the duration (of 8 days) was so similar in timescale and light-curve to J1957 that we decided to conduct a closer comparison with such objects.  

\subsection{Is J1957 an AMXP?}

The most extensive review of AMXPs is that of \citet{2012arXiv1206.2727P}, in which we found N6440 X-2 to be very similar to J1957 both in recurrence time ($\sim$1 month, the shortest recurrence time of any known XRT) and duration (3--5 d) of its outbursts \citep{Heinke2010}. The peak luminosities of the AMXPs are typically around 1\%$L_{Edd}$ ($\sim$10$^{36}$erg s$^{-1}$ for a $1.4M_\odot$ NS), which would place J1957 at $ d\sim 6\,\rm kpc$. Indeed, comparing directly with N6440 X-2 places J1957 at $d\sim 5\, \rm kpc$. Another key similarity here is the light-curve shape of at least two AMXP (SAX J1808.4-3658 and IGR J00291+5934) during their (short, 1--2 weeks) periods of activity (see Fig.~\ref{fig: zoom}). They are characterised by an initial slow decay from the peak luminosity that ends with a sharp drop (see \citealt{2008ApJ...675.1468H}, \citealt{2011ApJ...726...26H}). The knee of the light curve at a critical flux has been interpreted either as centrifugal inhibition of accretion or as the propagation of a cooling wave in the disc \citep{1998A&A...338L..83G}. Assuming that these phenomena take place at the same luminosity level, we infer a distance to J1957 of $d \sim 5-6\, \rm kpc$ (using $d=3.5\, \rm kpc$ for SAX J1808.4-3658 and $d=4\, \rm kpc$ for IGR J00291+5934). 

There are three ultracompact AMXPs ($P_{\rm orb}<1.3\, \rm{h}$) known to have exhibited short (below 2 weeks) outbursts: N6440 X-2 \citep{Altamirano2010}, XTE J1751-305 (\citealt{Grebenev2005}, \citealt{2012arXiv1206.2727P}) and Swift J1756.9-2508 \citep{Patruno2010}; which favours the association of short outbursts with short orbital period LMXBs. Nevertheless, we note that at least 3 other AMXPs with longer periods that have exhibited similarly short outbursts:: IGR J00291+5934 (\citealt{Hartman2011}; $P_{\rm orb}=2.5\, \rm{h}$ reported by \citealt{Markwardt2004}), Swift J1749.4-2807 ($P_{\rm orb}=8.82\, \rm{h}$, see \citealt{Wijnands2009} and \citealt{Altamirano2011}) and SAX J1748.9-2921 ($P_{\rm orb}=8.77\, \rm{h}$, \citealt{Patruno2009}). On the other hand, there are short period AMXPs like XTE J0929-314 \citep{Remillard2002} and XTE J1807-294 \citep{Markwardt2003} which exhibited outbursts lasting several months. Despite the fact that these examples highlight the inexactness of the correlation between short outbursts and short orbital periods, we propose J1957 as a short period LMXB due to the combination of short duration outbursts, high recurrence time and featureless optical spectrum.

Finally, assuming that the optical emission during active periods is dominated by an X-ray irradiated accretion disc, then we can use the X-ray/optical correlation equation of \cite{1994A&A...290..133V}: $ M_V = 1.57 - 2.27\, \log\Sigma$, where $\Sigma = (L_{\rm X}/L_{\rm Edd})^{1/2} (P_{\rm orb}[h])^{2/3}$ to derive an $M_V$ for J1957 of 3.4, assuming a 2h period. With our brightest B magnitude (see Sec. \ref{observations}), this would imply a distance of about $\sim 7-9\, \rm kpc$, somewhat closer if the period were shorter.

\subsection{The X-ray spectrum}
\label{xraydisc}

\begin{figure}
\scalebox{1.0}{\includegraphics[keepaspectratio, trim=0.cm 0.5cm 0.5cm 0.5cm, width=\columnwidth]{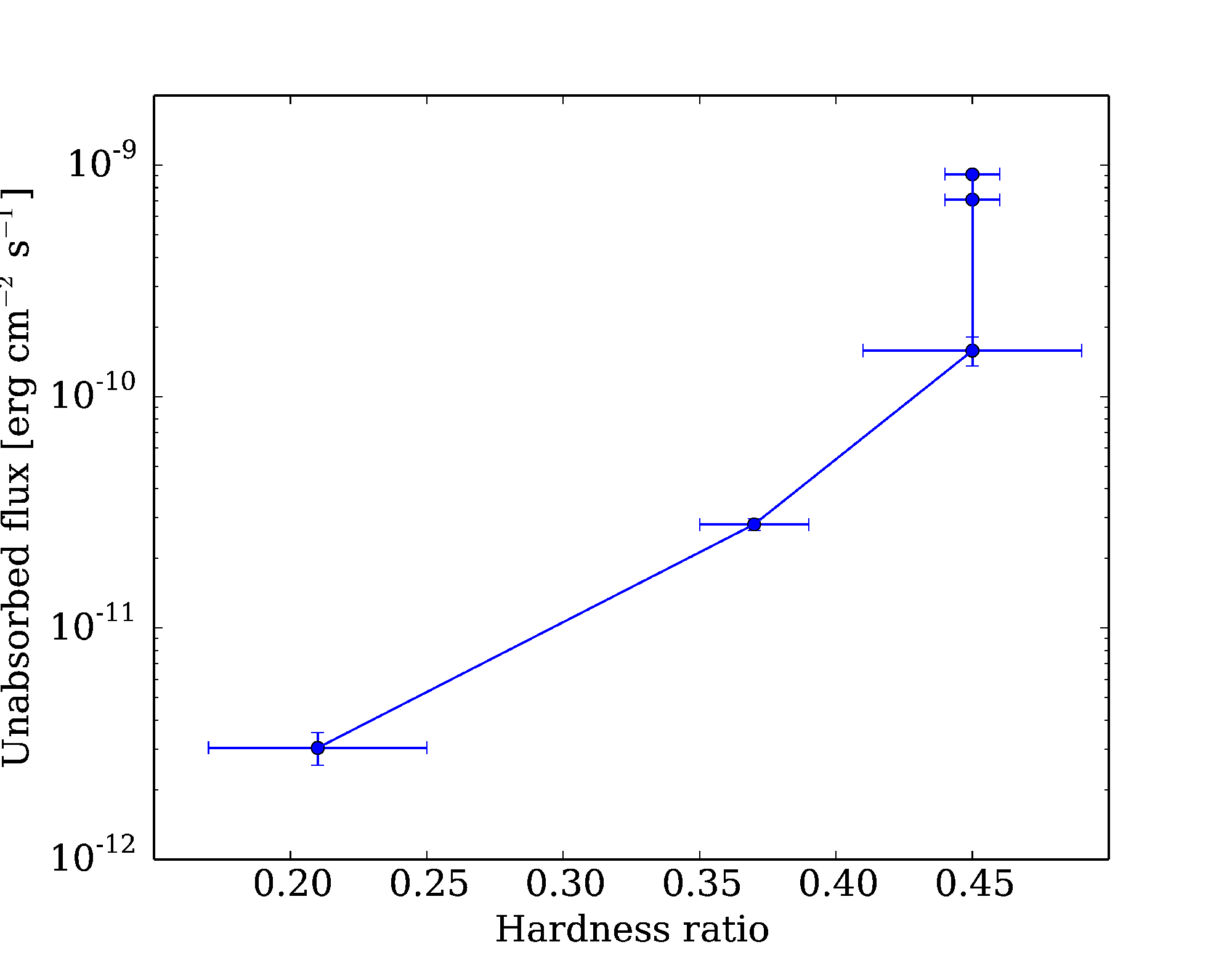}}
    \caption{Hardness ratio for the XRT detections in the last outburst, defined as the ratio between (2-10 keV) and (0.5-2 keV) fluxes. The largest error point corresponds to the observation affected by pile-up (see Sec. \ref{observations}).}
    \label{fig:hratio}
\end{figure}

Hysteresis patterns between Compton-dominated and thermal-dominated states have been observed in both BH and NS LMXBs (e.g. \citealt{Belloni2011}; \citealt{2014MNRAS.443.3270M}). These hysteresis phases, which are not witnessed in J1957, are observed at luminosities larger than $\sim 2\%\, L_{\rm edd}$ (see also \citealt{2003A&A...409..697M}). On the other hand, a softening phase during the decay towards quiescence is detected at lower luminosities (\citealt{2013ApJ...773...59P}, \citealt{2015MNRAS.454.1371W}). In J1957 we do observe a softening in the hardness ratio as the luminosity decreases by, at least, a factor of $\sim 50$ (see fig.~\ref{fig:hratio}). 

 Indeed, if the softening of the source begins at $1.6 \times 10^{-10}\rm erg\, cm^{-2} s^{-1}$, given that \citet{2015MNRAS.454.1371W} found this event always below $3 \times 10^{36}\rm erg\, s^{-1}$, which would place J1957 at $d< 13\,\rm kpc$. On the other hand, the non-detection in quiescence of J1957 in the USNO-B catalogue \citep{Monet2003}, whose limiting magnitude in V is $\sim 21$, implies a drop in luminosity of more than three magnitudes in the optical. If the donor is a dwarf star at a distance $d<13\,\rm kpc$, its absolute magnitude is constrained to $M_{\rm V} >5.4$. Using \citet{Cox2000} tabulated values, this results in a spectral type later than G7 (K6 if $d\sim 5\,\rm kpc$).

\section{Conclusions}

Our X-ray and optical analysis suggests that MAXI J1957+032 is a short period LMXB, given the short duration and  frequent recurrence of its outbursts, as well as the featureless optical spectrum. The spectral analysis leads us to place the system at $d<13\, \rm kpc$, which combined with the non-detection of the quiescence counterpart results in a donor star spectral type later than G8. After comparing with the properties of other known XRTs, we find similarities with those of the AMXP class, which would place J1957 at $d\sim 5\, \rm kpc$. Further multi-wavelength observations are encouraged especially taking into account the short recurrence times this system has exhibited in the past $\sim 1.5$ years.

\section*{Acknowledgements}

DMS acknowledges Fundaci\'on La Caixa for the financial support received in the form of a PhD contract. PAC is grateful for the support of a Severo Ochoa Visiting Research Fellowship and the hospitality of the Instituto de Astrof\'{i}sica de Canarias, where the work reported here was undertaken. We also acknowledge support by the Spanish Ministerio de Econom\'ia y competitividad (MINECO) under grant AYA2013-42627. TMD acknowledges support via a Ram\'{o}n y Cajal Fellowship (RYC-2015-18148). DAHB acknowledges support by the National Research Foundation and also thanks M. Kotze for supporting the SALT observations. MOLLY software developed by T. R. Marsh is gratefully acknowledged. This research has made use of \textit{MAXI} data provided by RIKEN, JAXA and the \textit{MAXI} team. We also acknowledge the use of public data from the \textit{Swift} data archive. Some of the observations reported in this paper were obtained with the Southern African Large Telescope (SALT).

%%%%%%%%%%%%%%%%%%%%%%%%%%%%%%%%%%%%%%%%%%%%%%%%%%



\bibliographystyle{mnras}
\bibliography{MAXI_J1957}

%\begin{thebibliography}{99}
%\end{thebibliography}

%%%%%%%%%%%%%%%%%%%%%%%%%%%%%%%%%%%%%%%%%%%%%%%%%%

% Don't change these lines
\bsp	% typesetting comment
\label{lastpage}
\end{document}